\documentclass[twocolumn,superscriptaddress,prl,showpacs,amsmath,amssymb,floatfix]{revtex4}
\usepackage{graphicx}\usepackage[bookmarks=false,colorlinks=true]{hyperref}

\begin{document}

\title{Persistence of locality in systems with power-law interactions}

\author{Zhe-Xuan Gong}
\email[The two authors contributed equally.]{}
\affiliation{Joint Quantum Institute, NIST/University of Maryland, College Park, Maryland 20742, USA}

\author{Michael Foss-Feig}
\email[The two authors contributed equally.]{}
\affiliation{Joint Quantum Institute, NIST/University of Maryland, College Park, Maryland 20742, USA}

\author{Spyridon Michalakis}
\affiliation{Institute for Quantum Information \& Matter, California Institute of Technology, Pasadena, California 91125, USA}

\author{Alexey V. Gorshkov}
\affiliation{Joint Quantum Institute, NIST/University of Maryland, College Park, Maryland 20742, USA}

\date{\today}

\begin{abstract}
Motivated by recent experiments with ultra-cold matter, we derive a new bound on the propagation of information in $D$-dimensional lattice models exhibiting $1/r^{\alpha}$ interactions with $\alpha>D$. The bound contains two terms: One accounts for the short-ranged part of the interactions, giving rise to a bounded velocity and reflecting the persistence of locality out to intermediate distances, while the other contributes a power-law decay at longer distances. We demonstrate that these two contributions not only bound but, except at long times, \emph{qualitatively reproduce} the short- and long-distance dynamical behavior following a local quench in an $XY$ chain and a transverse-field Ising chain. In addition to describing dynamics in numerous intractable long-range interacting lattice models, our results can be experimentally verified in a variety of ultracold-atomic and solid-state systems. 
\end{abstract}


\maketitle

In relativistic quantum theory, information propagation is limited by the speed of light. While the speed of light plays no such role in non-relativistic many-body quantum systems \cite{lieb72,nachtergaele10}, bounds on information propagation can emerge when interactions are short-ranged, as first shown by Lieb and Robinson \cite{lieb72}. Such Lieb-Robinson bounds underly our understanding of numerous equilibrium and non-equilibrium phenomena, including the generation of entanglement and topological order \cite{fukuhara13,langen13,polkovnikov11,bravyi06,nachtergaele_muchado_2011}, exponential decay of correlations in gapped ground states \cite{hastings05,nachtergaele_lieb-robinson_2006}, and the Lieb-Schultz-Mattis theorem \cite{hastings04,nachtergaele_multi-dimensional_2007}. While these bounds are well established, their generalization to systems with long-range interactions is far from complete \cite{hastings05,hauke13,eisert13}. Meanwhile, numerous currently available atomic, molecular, and optical systems exhibiting long-range interactions are emerging as versatile platforms for studying quantum many-body physics both in and out of equilibrium. These long-range interactions include dipolar $(1/r^{3})$ interactions between electric \cite{saffman10,yan13} or magnetic \cite{aikawa12,lu12,childress06,balasubramanian09,weber10,dolde13} dipoles, strong van-der-Waals $(1/r^{6})$ interactions between Rydberg atoms \cite{saffman10,schauss12} or polaritons \cite{firstenberg13}, along with $1/r^{\alpha}$ and even more general forms of interactions between trapped ions \cite{islam13,britton12,Richerme14,Jurcevic14} or atoms in multimode cavities \cite{gopalakrishnan11}.

One important consequence of a speed limit for short-range interacting systems is the existence of a linear light cone, which bounds a causal region and gives rise to a notion of locality. While Lieb-Robinson bounds have been generalized to long-range interacting systems by Hastings and Koma \cite{hastings05}, it is not yet clear to what extent locality persists. For example, while the Hastings-Koma bound allows for a causal region that grows exponentially in time, and thus a divergent velocity, to the best of our knowledge there are no models that explicitly demonstrate such behavior. Conversely, linear light cones have been observed in systems with long-ranged interactions \cite{hauke13,eisert13}, yet are manifestly absent in the existing Hastings-Koma bound.

\begin{figure}[t]
\includegraphics[width=1\columnwidth]{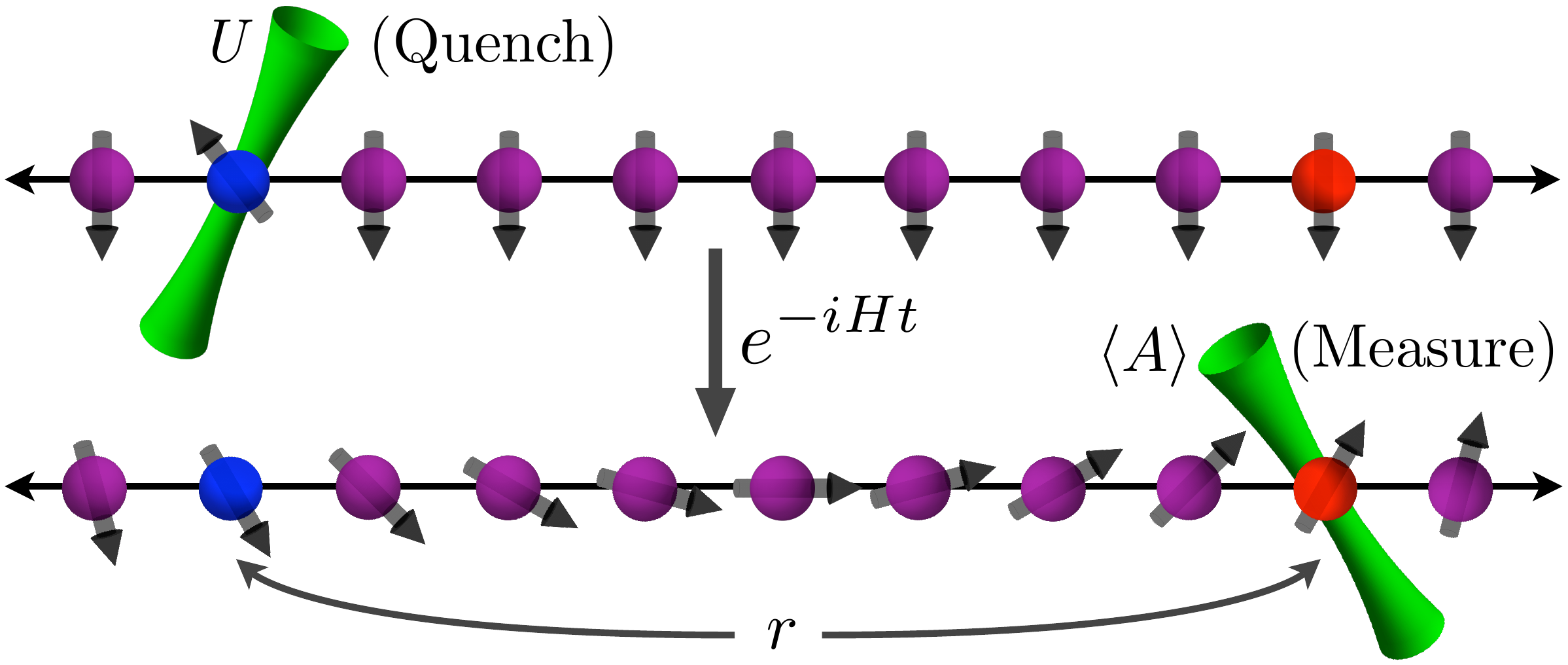} \vspace{-0.5cm}
\caption{(Color online). Illustration of the quench observable $Q_{r}(t)$. At time $t=0$, one spin (shown in blue) is perturbed by a unitary operator $U$. The effect on another spin (shown in red) a distance $r$ away, charecterized by the expectation value of an operator $A$, is measured at a later time $t$.} \label{fig:1}
\end{figure}

In this Letter, we study dynamics following a local quantum quench in spin systems with power-law ($1/r^{\alpha}$) interactions. To characterize the propagation of information, we consider the measurable quantity
\begin{equation}
Q_{r}(t)=\left|\langle\psi| U^{\dagger}A(t)U|\psi\rangle-\langle\psi| A(t)|\psi\rangle\right|/2. \label{eq:Qrt}
\end{equation}
Here $|\psi\rangle$ is an arbitrary initial state, $U$ is a unitary operator perturbing a single spin, while $A$ is an observable on a lattice site a distance $r$ away, and measured at a time $t$ later \cite{footnote0}. For convenience, we assume that the expectation value of $A$ in any state is between $-1$ and $1$, and hence $0\leq Q_r(t) \leq 1$.  Being the difference between the expectation values of $A(t)$ with and without the quench $U$, $Q_r(t)$ quantifies the ability to send information over a distance $r$ in a time $t$ (Fig.\,\ref{fig:1}).

In interacting many-body systems, it is not generally possible to calculate $Q_{r}(t)$ exactly; therefore, rigorous bounds on its behavior are crucial for characterizing how information propagates.  In what follows, we derive a  bound on $Q_{r}(t)$ that for the first time captures two crucial features of long-range interacting systems: (1) The persistence of a linear light cone at intermediate distances, and (2) The contraction of the causal region for decreasing interaction range, such that locality is recovered in the large-$\alpha$ limit. These features are then verified in the exact dynamics of two non-integrable lattice spin models. We note that our results can also be used to bound, for example, entanglement and correlation growth after a global quench \cite{Eisert10}, or ground-state correlations of gapped Hamiltonians \cite{hastings05}. Our results are relevant to recent experiments in trapped ions \cite{Richerme14,Jurcevic14}, and build on recent theoretical work studying post-quench dynamics in several long-range interacting systems, including Ising models with \cite{hauke13,schachenmayer13,knap13} and without \cite{foss-feig13,eisert13} a transverse field, the $XXZ$ chain \cite{hazzard13,eisert13}, and spin models with boson-mediated interactions \cite{juenemann13}.

\begin{figure}[t]
\includegraphics[width=1\columnwidth]{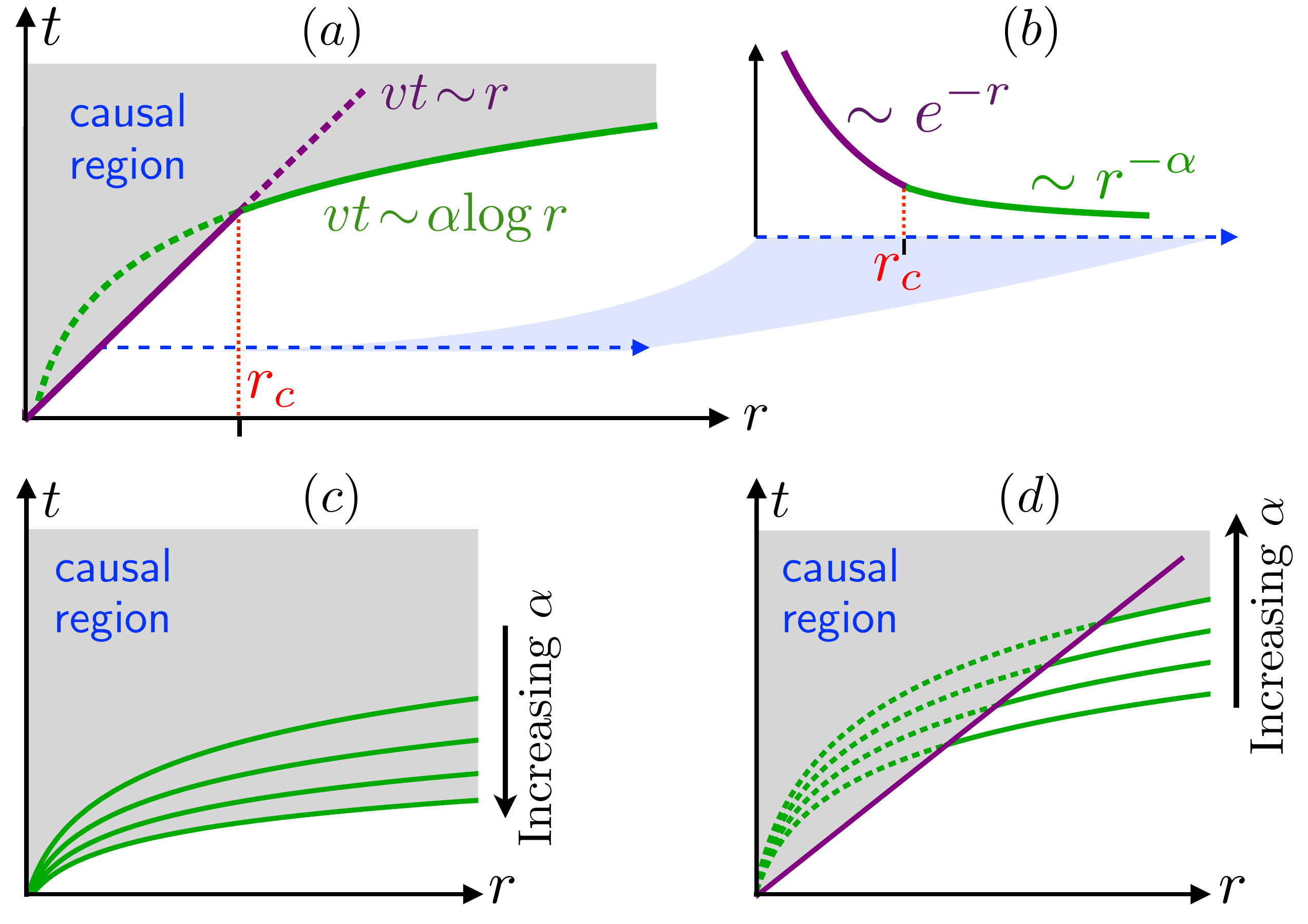} \vspace{-0.4cm} 
\caption{(Color online). (a) Illustration of the causal region (shaded) resulting from Eq.\,\eqref{eq:Bound} for the case $v_{1}=v_{2}=v$ and $\mu=1/2$. The boundary switches from linear to logarithmic at a critical $r_{c}$ satisfying $r_{c}\sim\alpha\log r_{c}$. (b) The decay of the signal outside the causal region changes from exponential to algebraic at $r_{c}$. (c) The Hastings-Koma bound \cite{hastings05} leads to a logarithmically bounded causal region that \emph{expands} with decreasing interaction range (increasing $\alpha$). (d) The new bound in Eq.\,\eqref{eq:Bound} gives rise to a causal region that \emph{contracts} with decreasing interaction range, converging to a linear light-cone for $\alpha\rightarrow\infty$.} \label{fig:2}
\end{figure}

\emph{Model and main results.---}For clarity of presentation, we consider a one-dimensional (1D) lattice, but generalizations to $D>1$ are straightforward. We study the post-quench dynamics of a spin model with Hamiltonian $H=\sum_{i<j}J_{ij}h_{ij}$ (so $A(t)=e^{iHt}Ae^{-iHt}$). Here the interaction $h_{ij}$ is a Hermitian operator acting on sites $i$ and $j$, whose expectation value in any state is between $-1$ and $1$.  The coupling constants are given by $J_{ij}=1/r_{ij}^{\alpha}$, where $r_{ij}=|i-j|$ (for convenience we set $J_{ii}=1$). In what follows, we will prove that 
\begin{equation}
Q_{r}(t)\leq c_{1}(e^{v_{1}t}-1)e^{-\mu r}+c_{2}(e^{v_{2}t}-1)/[(1-\mu)r]^{\alpha},\label{eq:Bound}
\end{equation}
The constants $c_{1},c_{2},v_{1},v_{2}$ are finite for all $\alpha>1$ and independent of $t$ and $r$, while $0<\mu<1$ is an adjustable parameter that can be tuned to optimize the bound for any particular value of $\alpha$.  Importantly, the bound does not depend on the form of the interaction $h_{ij}$, and therefore is applicable in many situations where exact (analytical or numerical) calculation of $Q_r(t)$ is not feasible. As shown in Fig.\,\ref{fig:2}(a), we can define a causal region as the part of the $r$-$t$ plane where the right-hand-side of Eq.\,\eqref{eq:Bound} is larger than a given value. The first term on the right-hand-side of Eq.\,\eqref{eq:Bound} is reminiscent of the familiar Lieb-Robinson bound \cite{lieb72}; alone it would lead to a causal region bounded by a linear light cone ($v_{1}t\gtrsim r$), and thus to a finite velocity for the propagation of information \cite{footnote1}. The second term is superficially similar to the Hastings-Koma bound \cite{hastings05}; alone it would lead to a causal region with a logarithmic boundary $v_{2}t\gtrsim\alpha\log r$, and an actual velocity that grows exponentially in time. The two terms together give a hybrid boundary, which switches from linear to logarithmic behavior at a critical $r_{c}$ satisfying $r_{c}\sim\alpha\log r_{c}$. As shown in Fig.\,\ref{fig:2}(b), the decay of signal outside the causal region changes from exponential to polynomial at $r_{c}$.

We emphasize that, despite the superficial similarity between the long-range piece of Eq.\,\eqref{eq:Bound} and the Hastings-Koma bound, they are fundamentally different, and Eq.\,\eqref{eq:Bound} \emph{cannot} be obtained by simply adding a short-range contribution to the Hastings-Koma bound. In order to not vanish (for all $r$ and $t$) in the large-$\alpha$ limit, where locality should be recovered, the Hastings-Koma bound requires a $v_{2}$ that diverges as $\alpha\rightarrow\infty$. Therefore, as shown in Fig.\,\ref{fig:2}(c), it leads to a causal region that actually grows \emph{larger} for shorter-range interactions. To the contrary, in the long-range piece of Eq.\,\eqref{eq:Bound}, $v_{2}$ remains finite in the large-$\alpha$ limit. Thus we obtain a much more physical scenario in which the causal region shrinks for progressively shorter-range interactions, eventually coinciding with the linear Lieb-Robinson light cone [Fig.\,\ref{fig:2}(d)].

\begin{figure}[t]
\centering \includegraphics[width=1\columnwidth]{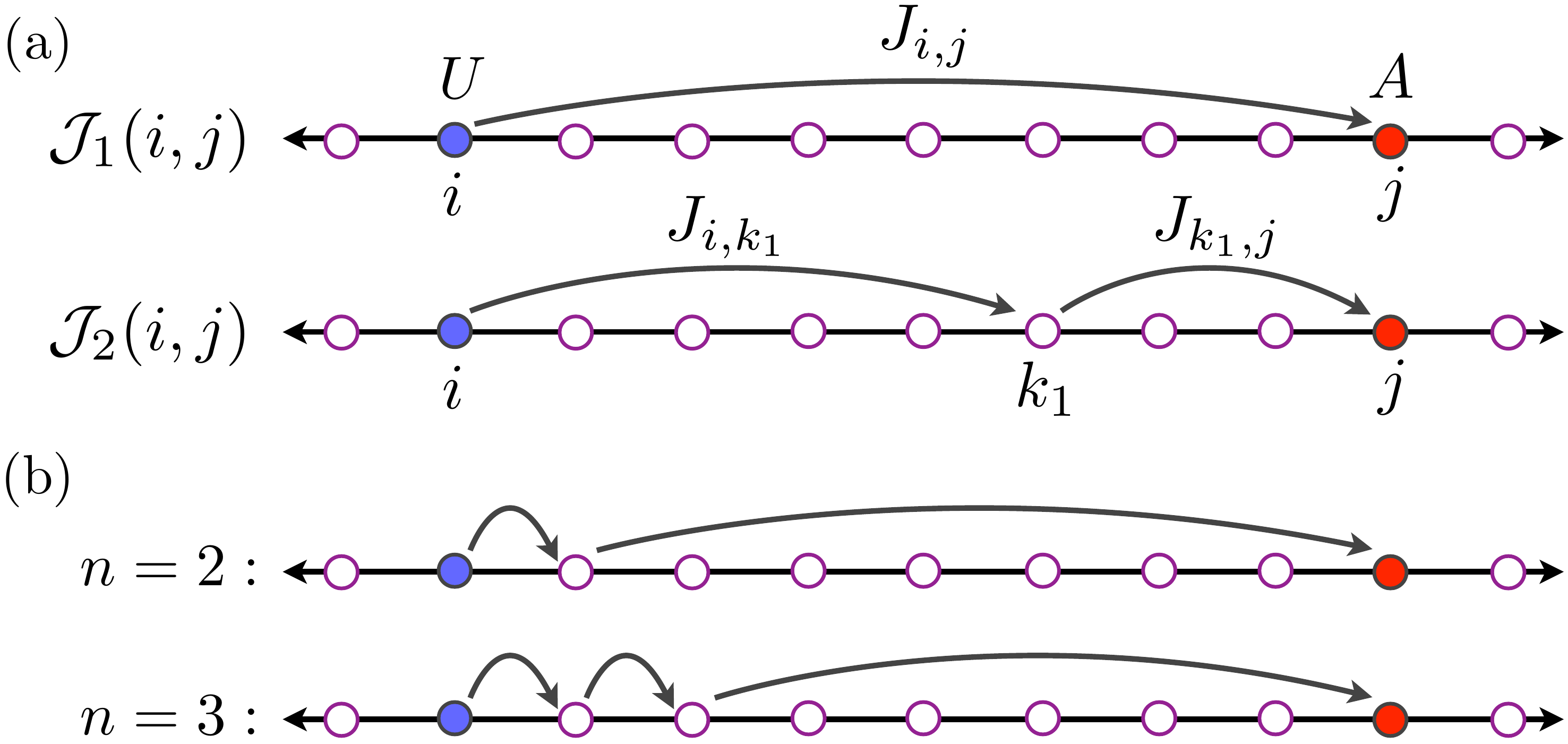}
\caption{(Color online) (a) Schematic representation of two hopping processes contributing to the first two terms in Eq.\,\eqref{eq:inf}. The full amplitude $\mathcal{J}_{n}(i,j)$ is obtained by summing the contribution from all possible $n^{{\rm th}}$ order paths. (b) The largest-magnitude terms contributing to the bound [Eq.\,\eqref{eq:newJbound}] on $\mathcal{J}_{n}(i,j)$ for $n=2,3$.} \label{fig:3} 
\end{figure}

\textit{Long-range generalization of Lieb-Robinson bounds.}---Using standard techniques \cite{lieb72,nachtergaele10,hastings05,juenemann13,WonderfulRussians}, $Q_{r}(t)$ can be bounded by an infinite series in time, 
\begin{eqnarray}
Q_{r}(t) & \leq & \sum_{n=1}^{\infty}\frac{(2\lambda t)^{n}}{n!}\mathcal{J}_{n}(i,j),\\ \label{eq:inf}
\mathcal{J}_{n}(i,j) & \equiv & \sum_{k_{1},\dots,k_{n-1}}J_{ik_{1}}J_{k_{1}k_{2}}\dots J_{k_{n-1}j}, \label{eq:Jn}
\end{eqnarray}
which has a clear physical interpretation. The quantity $\mathcal{J}_{n}(i,j)$ [depicted for $n=1,2$ in Fig.\,\ref{fig:3}(a)] can be thought of as the total contribution from all $n^{{\rm th}}$ order ``hopping'' processes connecting sites $i$ and $j$, with each ``hop'' being related to a matrix element of the Hamiltonian connecting the spins at its endpoints. The accompanying factor of $t^{n}$ arises simply because $t$ multiplies these matrix elements in the time-evolution operator. The factors of $\lambda\equiv\sum_{k}J_{ik}$ (which is a finite constant for all $\alpha>1$ because $\int_{1}^{\infty}dr/r^{\alpha}$ converges) are included for technical reasons to absorb otherwise present hops that originate from the initial (rather than final) site of the previous hop, thus allowing for the simple structure shown in Fig.\,\ref{fig:3}(a).

Physically, one would expect the total hopping amplitude $\mathcal{J}_{n}(i,j)$ to decay with $r\equiv r_{ij}$. This is manifestly true for $n=1$, where $\mathcal{J}_{1}(i,j)=J_{ij}=1/r^{\alpha}$. For the second-order hopping process, for any $i$, $j$, and $\alpha>1$ \cite{hastings05},  
\begin{equation}
\mathcal{J}_{2}(i,j)\equiv\sum_{k}J_{ik}J_{kj}\leq2\sum_{k}J_{ik}(2^{\alpha}J_{ij})=2\lambda2^{\alpha}J_{ij}. \label{eq:HK}
\end{equation}
The above inequality can be understood as follows: When the site $k$ is closer to $i$ than to $j$, the hop $J_{kj}$ must be longer than half of the distance between sites $i$ and $j$, and therefore corresponds to a hopping strength of at most $2^{\alpha}J_{ij}$.  The factor of $2$ then accounts for the terms in the sum when the site $k$ is closer to $j$ than to $i$.  Eq.\,\eqref{eq:HK} is called the \emph{reproducibility condition}, as repeated application of this inequality [to Eq.\,\eqref{eq:Jn}] gives $\mathcal{J}_{n}(i,j)\le(2\lambda2^{\alpha})^{n-1}/r^{\alpha}$, which reproduces the $1/r^{\alpha}$ decay for all $n$. Substituting these bounds on $\mathcal{J}_{n}(i,j)$ into Eq.\,\eqref{eq:inf} immediately yields the Hastings-Koma bound
\begin{equation}
Q_{r}(t)\leq c(e^{vt}-1)/r^{\alpha}, \label{eq:HKbound}
\end{equation}
where $v=4\lambda{}^{2}2^{\alpha}$ and $c=(2\lambda2^{\alpha})^{-1}$. The Hastings-Koma bound  \eqref{eq:HKbound} holds for all $\alpha>1$, so naively one would expect to be able to recover a short-ranged Lieb-Robinson bound [e.g. the first term in Eq.\,\eqref{eq:Bound}] by taking the limit $\alpha\rightarrow\infty$. However, because the velocity $v$ in Eq.\,\eqref{eq:HKbound} diverges exponentially with $\alpha$, the causal region actually encompasses all $r$ and $t$ for short-range ($\alpha\rightarrow\infty$) interactions [Fig.\,\ref{fig:2}(c)]. Below, we derive the new bound in Eq.\,\eqref{eq:Bound}, which recovers the correct short-range physics in the large-$\alpha$ limit, and, for finite $\alpha$, manifestly preserves the effects of short-range interactions at intermediate distance scales.

\textit{Recovering locality.}---To obtain the bound on $Q_{r}(t)$ given in Eq.\,\eqref{eq:Bound}, we begin by fixing the $2^{\alpha}$ divergence in the velocity $v$, which originates from the reproducibility condition Eq.\,\eqref{eq:HK}. The cause of this divergence is the attempt to bound repeated nearest-neighbor hops (which have unity amplitude for all $\alpha$) by a single long-range hop (whose amplitude decreases with $\alpha$). To resolve this issue, we separate out the nearest-neighbor hops in deriving the reproducibility condition in Eq.\,\eqref{eq:HK}:
\begin{equation}
\sum_{k}J_{ik}J_{kj}\le2(\sum_{r_{ik}\le1}J_{ik}J_{kj}+2^{\alpha}J_{ij}\sum_{r_{ik}\ge2}J_{ik}). \label{eq:modified_reproducability}
\end{equation}
Here the notation $\sum_{r_{ik}\le1}$ implies a sum over all sites $k$ for which $r_{ik}\le1$. Because the second sum $\sum_{r_{ik}\geq 2}J_{ik}$ does not contain nearest-neighbor hops, it can now be bounded by $2\sum_{r_{ik}~{\rm even}}r_{ik}^{-\alpha}=2\sum_{r_{ik}\geq   1}(2r_{ik})^{-\alpha}=2(\lambda-1)2^{-\alpha}$. Importantly, this cancels the $2^{\alpha}$ factor in Eq.\,\eqref{eq:modified_reproducability}.  Using the fact that $J_{ij}\leq\sum_{r_{ik}\leq 1}J_{ik}J_{kj}$, we then obtain
\begin{equation}
\sum_{k}J_{ik}J_{kj}\le4\lambda\sum_{r_{ik}\le1}J_{ik}J_{kj}.
\end{equation}
Applying this result iteratively in Eq.\,\eqref{eq:Jn}, we find
\begin{equation}
\mathcal{J}_{n}(i,j)\le(4\lambda)^{n-1}\!\!\!\!\!\!\!\!\!\!\!\!\!\!\!\!\sum_{r_{ik_{1}}\le1,\dots,r_{k_{n-2}k_{n-1}}\le1}\!\!\!\!\!\!\!\!\!\!\!\!\!\!\!\! J_{i,k_{1}}\dots J_{k_{n-1},j}. \label{eq:newJbound}
\end{equation}
The maximum possible value for each summand is given by $(r-n+1)^{-\alpha}$, corresponding to the hopping process containing $n-1$ nearest-neighbor hops from site $i$ towards $j$, together with one remaining hop of distance $r-n+1$ [see Fig.\,\ref{fig:3}(b)]. Since there are $3$ sites within unit distance of any given site,   $\mathcal{J}_{n}(i,j)\le(12\lambda)^{n-1}(r-n+1)^{-\alpha}$.

This new bound on $\mathcal{J}_{n}(i,j)$ is free of the $2^\alpha$ factor, as we desired. However, because $\mathcal{J}_{n}(i,j)$ now decays as $1/(r-n+1)^{\alpha}$ instead of the simple $1/r^{\alpha}$, the bound no longer decays with $r$ when $n$ approaches $r$ and fails when $n>r$. Thus to produce a useful bound, we must restrict the order $n$ to be smaller than some fraction of $r$. Denoting this fraction by a free parameter $\mu\in (0,1)$, to be optimized later, the contribution to $Q_r(t)$ of all hopping processes of order $n<\lceil \mu r \rceil$ (the smallest integer $\geq\mu r$) can be bounded as
\begin{equation}
\sum_{n=1}^{{\lceil \mu r \rceil}-1}\frac{(2\lambda t)^{n}}{n!}\mathcal{J}_{n}(i,j)\le c_{2}\frac{e^{v_{2}t}-1}{[(1-\mu)r]^{\alpha}}, \label{eq:long}
\end{equation}
with $v_{2}=24\lambda^{2}$ and $c_{2}=(12\lambda)^{-1}$.  Crucially, while this result superficially resembles the Hastings-Koma bound [Eq.\,\eqref{eq:HKbound}], the velocity $v_{2}$ no longer diverges with $\alpha$.

We still must bound the contribution from processes with $n\geq\lceil\mu r\rceil$ hops.  These processes are dominated by $\lceil\mu r\rceil$ repeated hops, each of a length $\sim1/\mu$.  Formally, 
\begin{equation}
\!\!\!\!\!\sum_{n={\lceil \mu r \rceil}}^{\infty}\!\!\!\frac{(2\lambda t)^{n}}{n!}\mathcal{J}_{n}(i,j)\le\!\!\!\sum_{n={\lceil \mu r \rceil}}^{\infty}\frac{(2\lambda^{2}t)^{n}}{n!\lambda e^{\mu r-n}} \!<\! c_{1}\frac{e^{v_{1}t}-1}{e^{\mu r}}, \label{eq:short}
\end{equation}
where we use the trivial bound $\mathcal{J}_{n}(i,j)\leq \lambda^{n-1}$. Here $v_{1}=2\lambda^{2}e$ and $c_{1}=\lambda^{-1}$. Not surprisingly, Eq.\,\eqref{eq:short} resembles the Lieb-Robinson bound, except that the exponential decay occurs on a typical length scale $1/\mu$.

Combining Eqs.\,\eqref{eq:long} and \eqref{eq:short}, we arrive at our bound in Eq.\,\eqref{eq:Bound}. A key feature of our new bound is that both velocities $v_{1}$ and $v_{2}$ actually \emph{decrease} (through the implicit $\alpha$-dependence of $\lambda$) with shorter interaction range (larger $\text{\ensuremath{\alpha}}$), consistent with the expected physical picture. Note that the free parameter $\mu$ can be optimized to give the best possible bound for a particular range of interactions. For example, in the  $\alpha\rightarrow\infty$ limit, the $1/r^\alpha$ part of our bound vanishes, and we can choose $\mu\rightarrow1$ to recover the Lieb-Robinson bound.  Alternatively, for small $\alpha$, choosing $\mu$ close to zero recovers the Hastings-Koma bound at sufficiently large $r$ (for small $\alpha$, the divergence with $\alpha$ of the Hastings-Koma velocity $v$ in Eq.\,\eqref{eq:HKbound} is not important). 

\textit{Applications to experimentally realizable models.}---We now show that the coexistence of behavior consistent with both terms in Eq.\,\eqref{eq:Bound} can be seen in experimentally realizable lattice spin models. We consider a spin-$1/2$ chain governed by (a) an $XY$ model: $H_{XY}=\frac{1}{2}\sum_{i<j}(\sigma_{i}^{x}\sigma_{j}^{x}+\sigma_{i}^{y}\sigma_{j}^{y})/r_{ij}^{\alpha}$, and (b) a transverse-field Ising model (TFIM): $H_{\mathrm{TFIM}}=\sum_{i< j}\sigma_{i}^{x}\sigma_{j}^{x}/r_{ij}^{\alpha}+B_{z}\sum_{i}\sigma_{i}^{z}$. Ions in a linear rf-Paul trap have already been used to simulate both models with $\alpha\in(0,3)$ \cite{islam13,Richerme14,Jurcevic14}. Alternately, for $\alpha=3$, both models can be simulated with polar molecules \cite{yan13,gorshkovPRA11,gorshkovPRL11,hazzard13}. In both models, we take a spin-polarized initial state $|\psi\rangle=\bigotimes_{i}|\sigma_{i}^{z}=-1\rangle$, apply a local quench operator $U=e^{i\pi\sigma_{0}^{y}/4}$ on site $0$, and measure $A=\sigma_r^x$ on site $r$.  For a chain with $N$ spins, we choose the time $1\lesssim t\ll N$ small enough to avoid boundary effects and large enough to prohibit a perturbative treatment of the dynamics.

\begin{figure}[t]
\includegraphics[width=1\columnwidth]{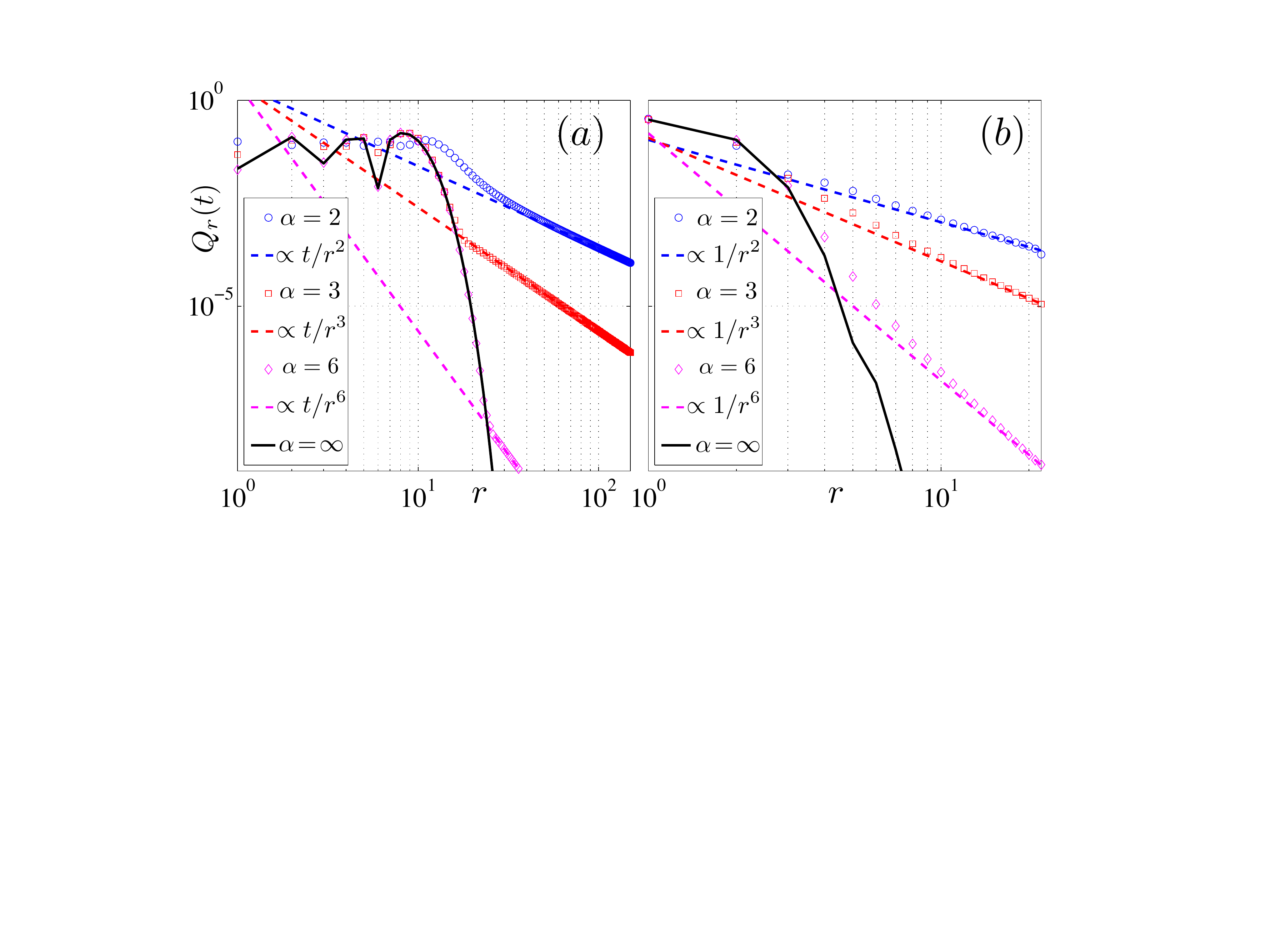}
\caption{(Color online). $Q_{r}(t)$ following a local quench in: (a) a $1/r^{\alpha}$ $XY$ chain (periodic boundary conditions, $N=501$, $t=5$), and (b) a $1/r^{\alpha}$ TFIM chain (open boundary conditions, $N=23$, $B_{z}=0.5$, $t=1$).} \label{fig:4}
\end{figure}

For the long-range $XY$ model subjected to the stated local quench, we can restrict our attention to the single spin-excitation subspace during the entire time evolution. As a result, we can map the spin model to a solitary free particle, making numerical calculation trivial for hundreds of spins. For $N=501$ spins and $\alpha=2,3,6,\infty$, Fig.\,\ref{fig:4}(a) demonstrates that at a specific time, the distance dependence of $Q_{r}(t)$ can be divided into several regions: (I) $1\le r\le r_{\mathrm{LC}}\equiv v_{\mathrm{max}}t$, where $v_{\mathrm{max}}$ denotes the maximum group velocity of the free particle. $Q_{r}(t)$ increases to its maximum value at $r\approx r_{\mathrm{LC}}$. (II) $r_{\mathrm{LC}}<r<r_{c}$, where $Q_{r}(t)$ decays faster than a power law. Note that for $\alpha=3$ and $\alpha=6$, $Q_{r}(t)$ is almost unchanged by the addition of long-ranged interactions for $r_{\rm LC}<r<r_{c}$. Thus the behavior of $Q_{r}(t)$ in this region is a direct consequence of nearest-neighbor interactions in the system, and is captured by the first term in Eq.\,\eqref{eq:Bound}. (III) $r>r_{c}$, where $Q_{r}(t)$ decays algebraically as $1/r^{\alpha}$ due to the second term in Eq.\,\eqref{eq:Bound}. Note, however, that $2Q_{r}(t)\approx t/r^{\alpha}$ (which is asymptotically exact in the limit of $t/r\rightarrow0$ \cite{Mike13}) does not saturate the time dependence $\exp(v_{2}t)-1$ in Eq.\,\eqref{eq:Bound}. This exponential time dependence in our bound (as well as in Ref.\,\cite{hastings05}) results from the $\mathcal{J}_{n}$ in Eq.\,\eqref{eq:inf} adding in phase. For the $XY$ model, a more careful analysis shows that the contributions for different $n$ do not add constructively, causing $Q_{r}(t)$ to depend linearly on $t$ \cite{Mike13}. These issues not-withstanding, it is abundantly clear (especially in the $\alpha=3$ and $\alpha=6$ cases) that the distance dependence of $Q_{r}(t)$ is a combination of a nearest-neighbor-interaction contribution (leading to rapid decay outside of a well-defined light cone) and a long-range-interaction contribution scaling as $1/r^{\alpha}$.

In the TFIM, the long-ranged interactions prevent a mapping onto a free model, and therefore our numerical calculation is limited to a relatively small chain size ($N=23$). Setting $B_{z}=0.5$, which accentuates the role of quantum fluctuations, we calculate $Q_{r}(t)$ numerically for $\alpha=2,3,6,\infty$ using a Krylov-subspace projection method. Figure \ref{fig:4}(b) shows that a local quench of the TFIM yields behavior that is qualitatively similar to the $XY$ model. For large $r$, we see a clear power-law decay $\sim1/r^{\alpha}$. For intermediate $r$, we see hints of faster than power-law decay similar to the nearest-neighbor case.

\textit{Outlook.}--- In addition to being relevant to a variety of equilibrium \cite{nachtergaele10,hastings04,hastings05,michalakis11b,michalakis12} and short-time non-equilibrium \cite{polkovnikov11,fukuhara13,langen13} phenomena, we also expect the derived bound to shed light on long-time relaxation processes in quantum many-body systems \cite{langen13}. It would be very interesting to try to either saturate or tighten the time dependence in the long-range part of the bound, thereby proving or ruling out the possibility of quantum state transfer \cite{bose07} in time $t\propto\log r$. 

We thank J.\,Preskill for asking whether the bound derived in Ref.\,\cite{hastings05} reduces to the nearest-neighbor case as $\alpha\rightarrow\infty$, and M. Kastner for pointing out that, if one optimizes with respect to $\mu$ at ever $r$ and $t$ in Eq.\,\eqref{eq:Bound}, the hybrid exponential-algebraic behavior shown in Fig.\,\ref{fig:4}  becomes evident only at larger values of $\alpha$. We thank A.\,M.\,Rey, K.\,Hazzard, C.\,Monroe, L.-M.\,Duan, C.\,Senko, P.\,Richerme, M.\,Maghrebi, A.\,Daley, J.\,Schachenmayer, A.\,Lee, J.\,Smith, and S.\,Manmana for discussions. This work was supported by the JQI and the NSF PFC at JQI. MFF thanks the NRC for support. SM acknowledges funding provided by the Institute for Quantum Information and Matter, an NSF Physics Frontiers Center with support of the Gordon and Betty Moore Foundation through grant GBMF1250.

\bibliographystyle{apsrev}
\bibliography{refs}

\begin{thebibliography}{43}
\bibitem[{foo({\natexlab{a}})}]{footnote0}
\bibinfo{howpublished}{If $U$ and $A$ are supported on general sets $X$ and
  $Y$, then the first and second terms in Eq.\,\eqref{eq:Qrt} become
  proportional to $\textrm{min}(|X|,|Y|)$ \cite{lieb72} and $|X||Y|$
  \cite{hastings05}, respectively.}
\bibitem[{foo({\natexlab{b}})}]{footnote1}
\bibinfo{howpublished}{For the sake of the qualitative discussion in this
  paragraph, the precise value of $\mu$ is unimportant, as long as both $\mu$
  and $1-\mu$ are on the order of $1$.}
\end{thebibliography}

\end{document}